\documentclass[aps,prd,showpacs,twocolumn,10pt,superscriptaddress,floatfix,nofootinbib,longbibliography]{revtex4-1}  
\usepackage{amssymb,amsfonts,amsmath,bm,bbm}   
\usepackage{graphicx}% Include figure files
\usepackage{dcolumn}% Align table columns on decimal point
\usepackage{bm}% bold math
\usepackage{CJK}
\usepackage{minted}  %%%%
\usepackage[%dvipdfm,
colorlinks=true,
linkcolor=blue,%black,
breaklinks=true,
urlcolor=blue,
citecolor=blue]{hyperref}
\usepackage{bbold}
\usepackage{epstopdf}

\newcommand{\HNUST}{\affiliation{
Hunan Provincial Key Laboratory of Intelligent Sensors and Advanced Sensor Materials, \\ School of Physics and Electronics, Hunan University of Science and Technology, Xiangtan 411201, China}} 
 
\newcommand{\BU}{\affiliation{
School of Physics, Beihang University, Beijing 102206, China}}

\newcommand{\ITP}{\affiliation{
CAS Key Laboratory of Theoretical Physics,
            Institute of Theoretical Physics, \\Chinese Academy of Sciences,
            Beijing 100190, China}}

\newcommand{\UOC}{\affiliation{
Department of Physics and Astronomy "Ettore Majorana",
University of Catania, Via S. Sofia 64, I-95123 Catania, Italy}} 

\newcommand{\INFN}{\affiliation{
INFN-Sezione di Catania, Via S. Sofia 64, I-95123 Catania, Italy}}

\newcommand{\IMNU}{\affiliation{
College of Physics and Electronic Information, Inner Mongolia Normal University, Hohhot, 010022, China}}
      
\begin{document}

\title{QCD axions and domain walls in dense matter under compact stellar conditions
}

\author{Zhen-Yan Lu} 
\email%[Corresponding author:~]
{luzhenyan@hnust.edu.cn}
\HNUST

\author{Shu-Peng Wang}\email{wsp@mail.hnust.edu.cn}\HNUST

\author{Qi Lu}
\email{luqi0@buaa.edu.cn}\BU\ITP

\author{Bo-Nan Zhang}\email{zhangbn@lzu.edu.cn}\IMNU

\author{Marco Ruggieri}
\email{marco.ruggieri@dfa.unict.it}\UOC\INFN

\date{\today}

\begin{abstract}
In compact stellar environments, the stability of dense QCD matter requires the simultaneous fulfillment of charge neutrality and beta equilibrium. 
In this work, we study how temperature and finite chemical potential affect QCD topology and axion properties within this medium, analyzing both cases with and without the charge neutrality condition. 
Our results show that the topological susceptibility and axion properties are highly sensitive to the critical behavior of the chiral phase transition in both cases.
In particular, the axion mass is strongly suppressed near the transition, while the axion self-coupling constant develops a pronounced peak whose magnitude depends on the temperature and density of the medium. 
Remarkably, around the critical point at $T\simeq70$ MeV and $\mu\simeq346$ MeV, the self-coupling constant is enhanced by more than a factor of seven compared to its vacuum value, a feature that to the best of our knowledge has not been reported in previous studies. 
Such a strong amplification at the phase boundary indicates that axion-mediated interactions could play an important role in shaping the structure and stability of compact stars, with potential implications for their evolution and observable astrophysical signatures. 

\end{abstract}

\maketitle
%\tableofcontents

\section{Introduction} \label{sec:INTRODUCTION}

The quantum chromodynamics (QCD)~\cite{Gross:2022hyw} vacuum possesses non-trivial topological features that are encoded in CP-even topological cumulants~\cite{Guo:2015oxa,Kawaguchi:2023olk}. These properties arise from instanton configurations~\cite{Gross:1980br,tHooft:1986ooh,Schafer:1996wv}, which generate a $\theta$ term in the QCD Lagrangian. Although this term does not affect the classical dynamics, it induces a contribution to the electric dipole moment of neutral particles. Experimental measurements place an extremely stringent upper bound on the $\theta$ parameter~\cite{Baker:2006ts,Griffith:2009zz,Parker:2015yka,Graner:2016ses,Guo:2015tla}, leading to the long-standing strong CP problem. A compelling solution to this problem is provided by the Peccei–Quinn mechanism~\cite{Peccei:1977hh,Peccei:1977ur}, which predicts the existence of the QCD axion~\cite{Weinberg:1977ma,Wilczek:1977pj}, a pseudoscalar particle with profound implications for both particle physics and cosmology~\cite{Carenza:2024ehj,Marsh:2015xka}.

The QCD axion not only resolves the strong CP problem but also emerges as a well-motivated dark matter candidate, making it a subject of intense theoretical and experimental study~\cite{DiLuzio:2020wdo,Bradley:2003kg}. 
While direct detection remains elusive, axions may leave characteristic imprints in extreme astrophysical environments such as neutron stars and compact star mergers~\cite{Cao:2024cym,Hamaguchi:2025ztd}. 
These systems are governed by ultra-high densities and, in certain regimes, elevated temperatures, where exotic phases of quark matter may develop and axion production or absorption can significantly alter stellar dynamics~\cite{Cavan-Piton:2024ayu,Maruyama:2017xzl}. A quantitative understanding of axion behavior under such conditions requires a framework that incorporates both temperature and density effects, while consistently accounting for the interplay among chiral symmetry breaking, color superconductivity, and the nontrivial topological structure of QCD.

Beyond its role in particle physics, the axion has profound implications in astrophysics, where it may influence supernova explosions and protoneutron star evolution~\cite{Lucente:2020whw,Lucente:2022vuo,Fischer:2021jfm,Choi:2021ign}. They can be produced via the Primakoff process, where high-energy photons scatter off atomic nuclei to generate pseudoscalar particles~\cite{Masso:1995tw}. In addition, axion emission offers an efficient cooling mechanism for compact stars, acting as a complement to the conventional neutrino- and photon-driven cooling channels~\cite{Leinson:2014ioa,Sedrakian:2015krq,Sedrakian:2018kdm,Buschmann:2021juv,Buschmann:2019pfp}. 
Previous studies have mainly focused on axion emission and scattering processes in dense environments~\cite{DiLuzio:2024vzg,Springmann:2024mjp,Wang:2024tre}, as well as the coupling strength between axions and Standard Model particles~\cite{Vonk:2020zfh,Vonk:2021sit,Alonso-Alvarez:2018irt}. However, most of these works are restricted to zero-temperature conditions or to the couplings themselves, while investigations of the evolution of the axion mass and self-coupling constant at finite temperature and high density remain limited. %As noted in Ref.~\cite{Lu:2018ukl}, 
Chiral perturbation theory (CHPT) provides precise results for the $\theta$ vacuum topological susceptibility at zero temperature~\cite{GrillidiCortona:2015jxo,Lu:2020rhp}, but becomes unreliable at finite temperature and density~\cite{Carignano:2016rvs,Adhikari:2019zaj}, particularly near the chiral phase transition~\cite{Lu:2018ukl}, where it breaks down. On the other hand, although lattice QCD has made remarkable progress at zero baryon density, its extension to finite density is severely hindered by the notorious sign problem, which poses a major obstacle to obtaining reliable results at nonzero chemical potential and finite temperature~\cite{Splittorff:2007ck}. A comprehensive understanding of axion physics under realistic compact star conditions is therefore still lacking.

The NJL model~\cite{Nambu:1961fr,Nambu:1961tp,Hatsuda:1994pi,Klevansky:1992qe} provides a versatile framework for investigating the thermodynamic properties of strongly interacting quark matter and the phase transitions of QCD~\cite{Gross:2022hyw,Achenbach:2023pba,huston2023Quantum}. It successfully captures key qualitative features of the phase diagram, such as chiral symmetry restoration at high temperature, which lies beyond the range of applicability of CHPT~\cite{Gasser:1983yg}. The model has been shown to reproduce both zero-temperature~\cite{Avancini:2019ego,Lu:2019diy} and finite-temperature~\cite{Lopes:2021tro,Lu:2021hvw} results for isospin-imbalanced matter, in good agreement with lattice QCD and CHPT. In particular, it reproduces the characteristic peak structure of the energy density normalized to its Stefan–Boltzmann limit~\cite{Lu:2019diy}, consistent with CHPT predictions~\cite{Carignano:2016rvs} and lattice data~\cite{Detmold:2012wc}. It also correctly identifies the critical point of the transition from the normal phase to the pion superfluid phase at $\mu_I=m_\pi$~\cite{He:2005nk,Kogut:2002zg,Kogut:2004zg,Son:2000xc}. More recently, the axion field has been incorporated into the NJL model Lagrangian, first at finite temperature~\cite{Lu:2018ukl} and later at finite chemical potential~\cite{Zhang:2023lij,Gong:2024cwc}, enabling systematic studies of in-medium modifications of axion properties under conditions relevant to dense stellar environments. 
Building on these developments, the axion field is incorporated into the NJL Lagrangian through the $U(1)_A$ anomaly term, providing a direct and consistent approach to investigate axion dynamics in hot and/or dense QCD matter~\cite{Murgana:2025dfa,Zhang:2025lan,Carlomagno:2025ayh,Lopes:2022efy,Kumar:2024abb,Murgana:2024djt,Das:2020pjg,Horvatic:2019lok,Liu:2024spj,Chu:2024mmc,Bandyopadhyay:2019pml,zhang2025$theta$term}. 
In this work, we extend our previous study on QCD topology and axion properties to include the electric charge neutrality condition, which is essential for accurately modeling matter inside compact stars. While charge neutrality has been considered in earlier works~\cite{Zhang:2023lij}, our focus is distinct: we investigate the evolution of the lowest-order cumulants of the QCD topological charge distribution and their impact on the axion potential, mass, self-coupling, and domain wall tension as functions of chemical potential, with particular emphasis on low temperatures and high densities. These conditions closely resemble those inside compact star cores, making our results directly relevant to astrophysical applications. 
By analyzing axion properties in hot and dense quark matter with a model that treats scalar and pseudoscalar condensates in both quark–antiquark and diquark channels, we are able to capture the rich interplay of competing order parameters across different regions of the QCD phase diagram. This approach sheds light on the stability of axion-induced domain walls and their possible role in compact-star phenomenology, potentially affecting stellar cooling and energy transport. Our findings contribute both to the theoretical understanding of axions as dark matter candidates and to the broader effort of mapping the QCD phase structure under extreme conditions.

The structure of the paper is as follows. In Sec.~\ref{sec:NJLMODEL}, we provide a concise overview of the NJL model and its extension to include the axion field at finite temperature and chemical potential. Sec.~\ref{sec:results} is dedicated to the presentation of our numerical calculation results, with particular emphasis on the effects of temperature and chemical potential on domain wall tension, the first two cumulants of the QCD topological charge distribution, as well as on axion properties, both with and without the imposition of charge neutrality. 
Concluding remarks and a discussion of the implications of our results are given in Sec.~\ref{sec:CONCLUSION}.

%examined in both the real case and the chiral limit. 

%%%%%%%%%%%%%%%%%%%%%%%%%%%%%%%%%%%%%%%%
\section{NJL model} \label{sec:NJLMODEL}

To investigate the properties of the QCD axion in dense quark matter, we employ the two‑flavor NJL model as an effective framework, considering both charge-neutral and non-neutral scenarios relevant to compact stars. 
The corresponding Lagrangian density including the quark chemical potential current and the U(1)$_A$ symmetry-breaking term reads 
%~\cite{Buballa:2003qv,Hatsuda:1994pi,Klevansky:1992qe,Nambu:1961tp,Nambu:1961fr}
%\begin{eqnarray}
%   \mathcal{L}=\bar{q}\left(i \gamma+\hat{\mu} \gamma_0-m_0\right) q+\bar{e}\left(i \gamma+\mu_e \gamma_0\right) e+\mathcal{L}_{\text {int }} ,
%\end{eqnarray}
\begin{eqnarray} \label{eq:Lagrangian}
\begin{aligned}
\mathcal{L}= &~ \bar{q}\left(i \gamma^\mu\partial_\mu+\hat{\mu} \gamma_0-m_0\right) q+\bar{e}\left(i \gamma^\mu\partial_\mu+\mu_e \gamma_0\right) e \\
  % \mathcal{L}_{\text {int }}= 
   &+ G_1\left[\left(\bar{q} \tau_a q\right)\left(\bar{q} \tau_a q\right)+\left(\bar{q} \tau_a i \gamma_5 q\right)\left(\bar{q} \tau_a i \gamma_5 q\right)\right] \\
   & +8 G_2\left[e^{i \theta } \operatorname{det}\left(\bar{q}_R q_L\right)+e^{-i \theta } \operatorname{det}\left(\bar{q}_L q_R\right)\right] ,
\end{aligned}
\end{eqnarray}
where $q\equiv(u,d)^T$ denotes the quark field matrix, $e$ is the electron field, and $m_0$ is the current quark mass. The quark chemical potential matrix in flavor space is
$\hat{\mu}=\text{diag}\{\mu_u,~\mu_d\}$. 
The interaction term in the third line of Eq.~(\ref{eq:Lagrangian}) is adopted directly from 
Refs.~\cite{Lu:2018ukl}. 
In particular, the second line in the aforementioned equation represents the $U(1)_A$ symmetry-breaking term, which generates the coupling between the QCD axion and the quark fields. %~\cite{76Hooft3432-3450PRD,tHooft:1986ooh}. 
Here, $\tau_0$ denotes the identity matrix, while $\tau_i$ for $i=1,2,3$ correspond to the Pauli matrices. The interaction that preserves $U(1)_A$ symmetry is controlled by the coupling constant $G_1$, whereas $G_2$ governs the intensity of the term that violates $U(1)_A$ symmetry.

%After taking the mean-field approximation, 
Within the mean-field approximation, the thermodynamic potential at one loop can be expressed as %~\cite{Lu:2018ukl}
\begin{eqnarray}
   \Omega=\Omega_{\mathrm{mf}}+\Omega_e+\Omega_q ,
\end{eqnarray}
where the mean field contribution is given by 
\begin{eqnarray}\label{eq:OmegaAll}
\begin{aligned}
  \Omega_{\mathrm{mf}}= & -G_2\left(\eta^2-\sigma^2\right) \cos \theta +G_1\left(\eta^2+\sigma^2\right) \\
  & -2 G_2 \sigma \eta \sin \theta %,
\end{aligned}
\end{eqnarray}
with the scalar and pseudoscalar condensates defined as $\sigma=\langle\bar{q} q\rangle$ and $\eta=\left\langle\bar{q} i \gamma_5 q\right\rangle$. In addition, the thermodynamic potential density of electrons at finite temperature 
is given by 
\begin{equation}
\Omega_e=-\frac{\mu_Q^4}{12 \pi^2}-\frac{\mu_Q^2 T^2}{6}-\frac{7 \pi^2 T^4}{180}. 
\end{equation}
In the above, the electron mass has been neglected, as it is negligible compared to the quark masses and the relevant chemical potentials. Moreover, the contribution from the quark loop is given by
\begin{eqnarray}\label{eq:OmegaQ}
\begin{aligned}
  \Omega_q= & -2 N_c T \sum_{f=u, d} \int \frac{d^3 p}{(2 \pi)^3} 
\bigg\{\frac{E_p}{T} \\
&+
\ln 
\left[1+e^{-\left(E_p-\mu_f\right)/T}\right]
+\ln\left[1+e^{-\left(E_p+\mu_f\right)/T}\right]
\bigg\},
\end{aligned}
\end{eqnarray} 
with the dispersion laws of quarks are given by
\begin{eqnarray}
E_p=\sqrt{p^2+\Delta^2}, \quad \Delta^2=\left(m_0+\alpha_0\right)^2+\beta_0^2,
\end{eqnarray}
with
\begin{eqnarray}
\begin{aligned}
& \alpha_0=-2\left(G_1+G_2 \cos \theta \right) \sigma+2 G_2 \eta \sin \theta , \\
& \beta_0=-2\left(G_1-G_2 \cos \theta \right) \eta+2 G_2 \sigma \sin \theta  .
\end{aligned}
\end{eqnarray}

Considering the electroweak reactions in the quark star, $d \leftrightarrow u+e^{-}+\bar{v}_e$, 
we have to take into account the chemical equilibrium, which is perfectly achieved by 
\begin{eqnarray}
    \mu_u+\mu_e=\mu_d\equiv \mu. 
\end{eqnarray}
In this case, the chemical potentials of the up and down quarks, $\mu_u$ and $\mu_d$, can be equivalently expressed in terms of the baryon chemical potential $\mu$ and the charge chemical potential $\mu_Q$, namely 
\begin{align}
  & \mu_u=\mu-\frac{2}{3} \mu_Q, \label{eq:muU}\\%\quad 
  & \mu_d=\mu+\frac{1}{3} \mu_Q , \label{eq:muD}
\end{align}
where $\mu_Q$ denotes the electric chemical potential. 
For dense quark matter inside compact stars, the stability of the system requires the enforcement of electric charge neutrality, which is implemented by 
\begin{eqnarray}\label{eq:Q0}
\frac{\partial \Omega}{\partial \mu_Q}=0 .
\end{eqnarray}
Moreover, the condensates are computed self-consistently by solving the gap equations
\begin{eqnarray}\label{eq:gapEquations2}
\frac{\partial \Omega}{\partial \sigma}=0, \quad \frac{\partial \Omega}{\partial \eta}=0,
\end{eqnarray}
to ensure
that the solution $\sigma=\bar{\sigma}, \eta=\bar{\eta}$ indeed corresponds to the global minimum of the thermodynamic potential density of the system. Once the ground state is determined, the system is uniquely defined, and all relevant thermodynamic properties can be consistently derived from the thermodynamic potential evaluated at this minimum.

%\end{widetext}

%%%%%%%%%%%%%%%%%%%%%%%%%%%%%%%%%%%%%%%%
\section{Numerical results and discussions} \label{sec:results}

%%%%%%%%%已经改好
We notice that the first integral on the right-hand side of Eq.~(\ref{eq:OmegaQ}) is ultraviolet divergent, which we regularize by introducing a cutoff at $p=\Lambda$. In this case, the values of the NJL model parameters are taken as~\cite{Lu:2018ukl} $\Lambda=590~\mathrm{MeV}$, $G_0 \Lambda^2=2.435$, $G_1=(1-c) G_0$, $G_2=c G_0$, $c=0.2$, $m_0=6~\mathrm{MeV}$, which can be fixed by reproducing the empirical values of the pion mass $m_\pi=140.2$ MeV, the pion decay constant $f_\pi=92.6$ MeV, and the quark condensate in vacuum $\sigma_0=2(-241.5~\text{MeV})^3$. Thus, in principle for a given chemical potential $\mu$ we can solve numerically Eqs.~(\ref{eq:Q0}) and (\ref{eq:gapEquations2}) to obtain the quark and electron chemical potentials and the meson condensates, and thus all the thermodynamic
quantities of the system.

\begin{figure}[htb]  %bt
   \includegraphics[width=0.48\textwidth]{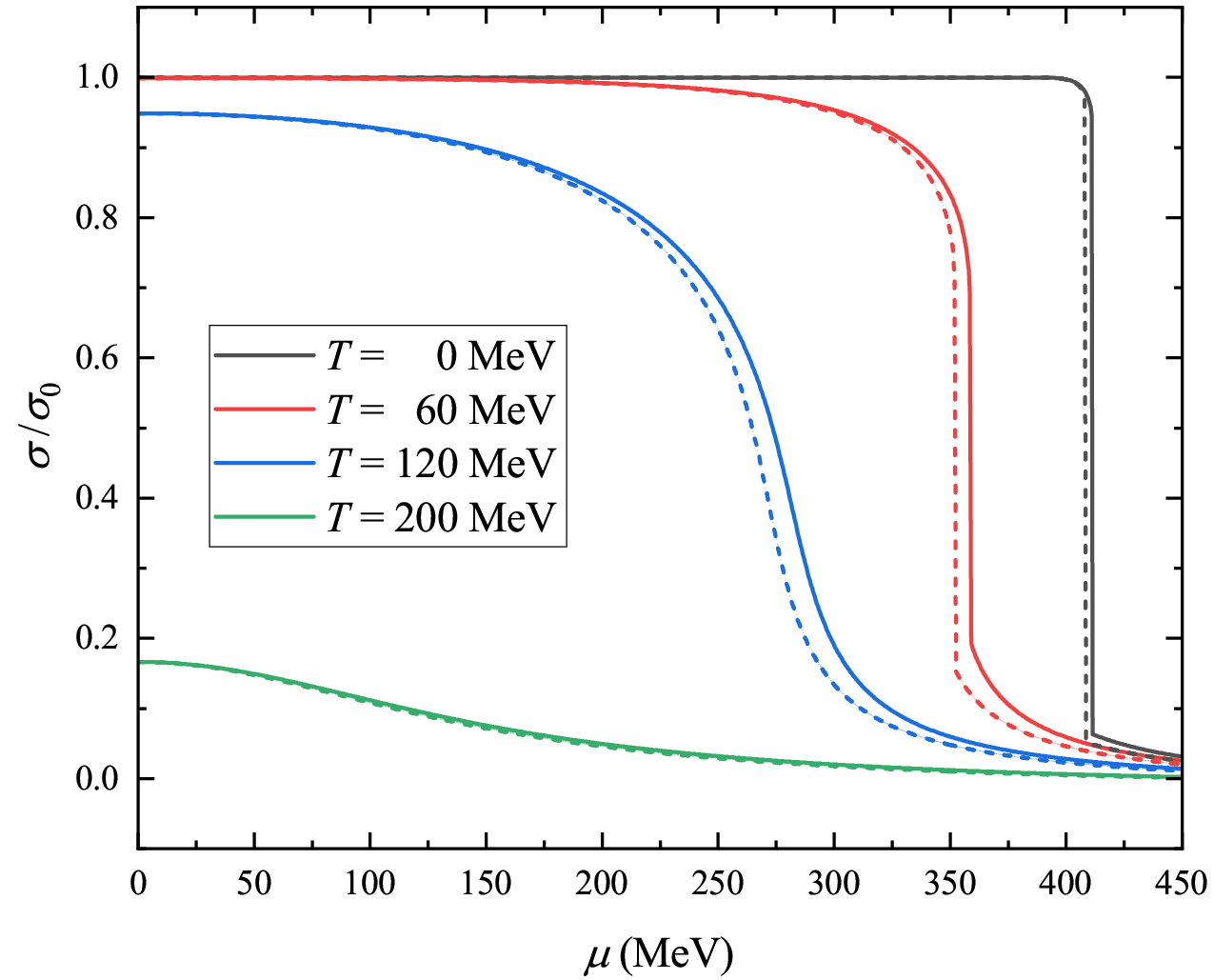}\\
  \caption{Chiral condensate, scaled by its value in the vacuum, as a function of the chemical potential at different temperatures with (solid) and without (dashed) electric charge neutrality. 
  }\label{fig:sigmamu}
\end{figure}

The spontaneous breaking and restoration of chiral symmetry is a fundamental aspect of QCD. The chiral condensate serves as an order parameter for this transition, with its suppression signaling the effective restoration of chiral symmetry at high temperature or density.
In Fig.~\ref{fig:sigmamu}, we plot the chiral condensate scaled by its vacuum value as a function of the chemical potential for several values of temperature: $T=0$ (black), $T=60$ MeV (red), $T=120$ MeV (blue), and $T=200$ MeV (green). 
Solid lines correspond to calculations with electric charge neutrality, while dashed lines represent results without this constraint. At $T=0$ and $60$ MeV, the condensate remains nearly unchanged at low chemical potential and then undergoes a sharp discontinuity at a critical value, reflecting a first-order chiral phase transition. At $T=120$ MeV, the condensate still decreases rapidly but the transition is smoother, indicating the weakening of the first-order character. At $T=200$ MeV, the condensate falls gradually with chemical potential, consistent with a smooth crossover. The comparison between the solid and dashed curves shows that enforcing charge neutrality and beta equilibrium does not qualitatively modify the condensate profile.
The strong dependence of the condensate on both temperature and chemical potential underscores its central role in the thermodynamics of dense QCD matter. As will be showed below, this behavior also influences other observables, including the axion mass and self-coupling, thereby linking chiral dynamics with axion phenomenology in compact stars.

%%%%%%%%%%%%%%%
\subsection{Axion potential and surface tension of the axion domain wall}

\begin{figure}[htb] 
   \includegraphics[width=0.48\textwidth]{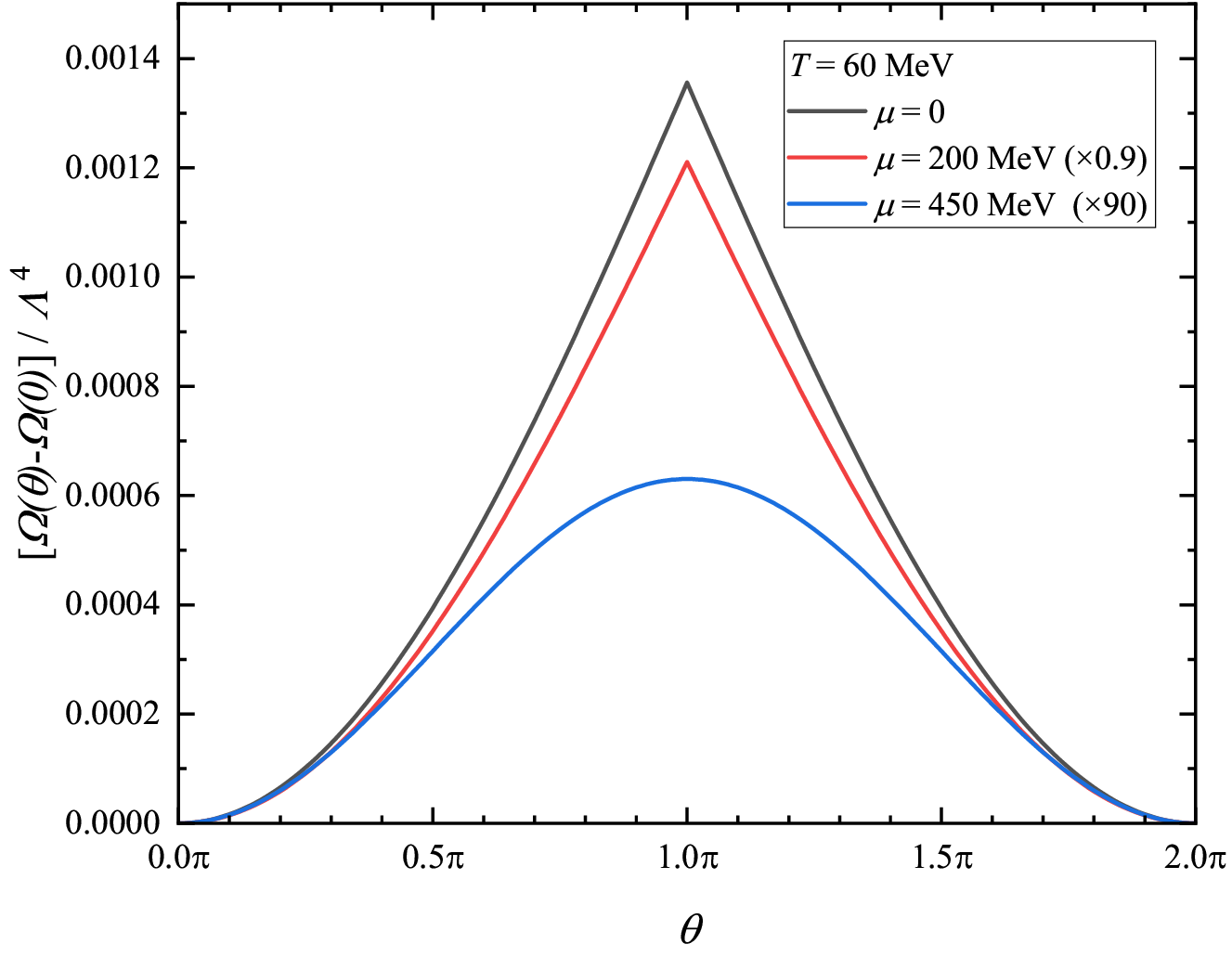}\\
  \caption{Thermodynamic potential as a function of $\theta$ at different chemical potentials. The potential is measured in units of the momentum cutoff $\Lambda$.
  %Conventions for colors  are the same used in Fig.~\ref{fig:sigmamu}. 
  }\label{fig:OmigaThetaT60}
\end{figure}

In Fig.~\ref{fig:OmigaThetaT60}, we show the axion effective potential as a function of the vacuum angle $\theta$ at $T=60$ MeV for chemical potentials $\mu=0$, 200, and 450 MeV. The overall shape exhibits the expected periodicity, with a maximum at $\theta=\pi$. As the chemical potential increases, both the height and sharpness of the peak are reduced, resulting in a progressively broader and flatter profile. This trend reflects the suppression of the topological susceptibility at finite density, in close analogy with the effect of increasing temperature. Physically, the reduction of the potential barrier indicates that the distinction between neighboring $\theta$ vacua becomes less pronounced in the chiral-restored phase. Consequently, the energy cost of forming axion domain walls is significantly lowered, which could strongly affect their stability and dynamical evolution in dense QCD matter. %~\cite{Zhang:2023lij}.

\begin{figure}[htb] 
   \includegraphics[width=0.48\textwidth]{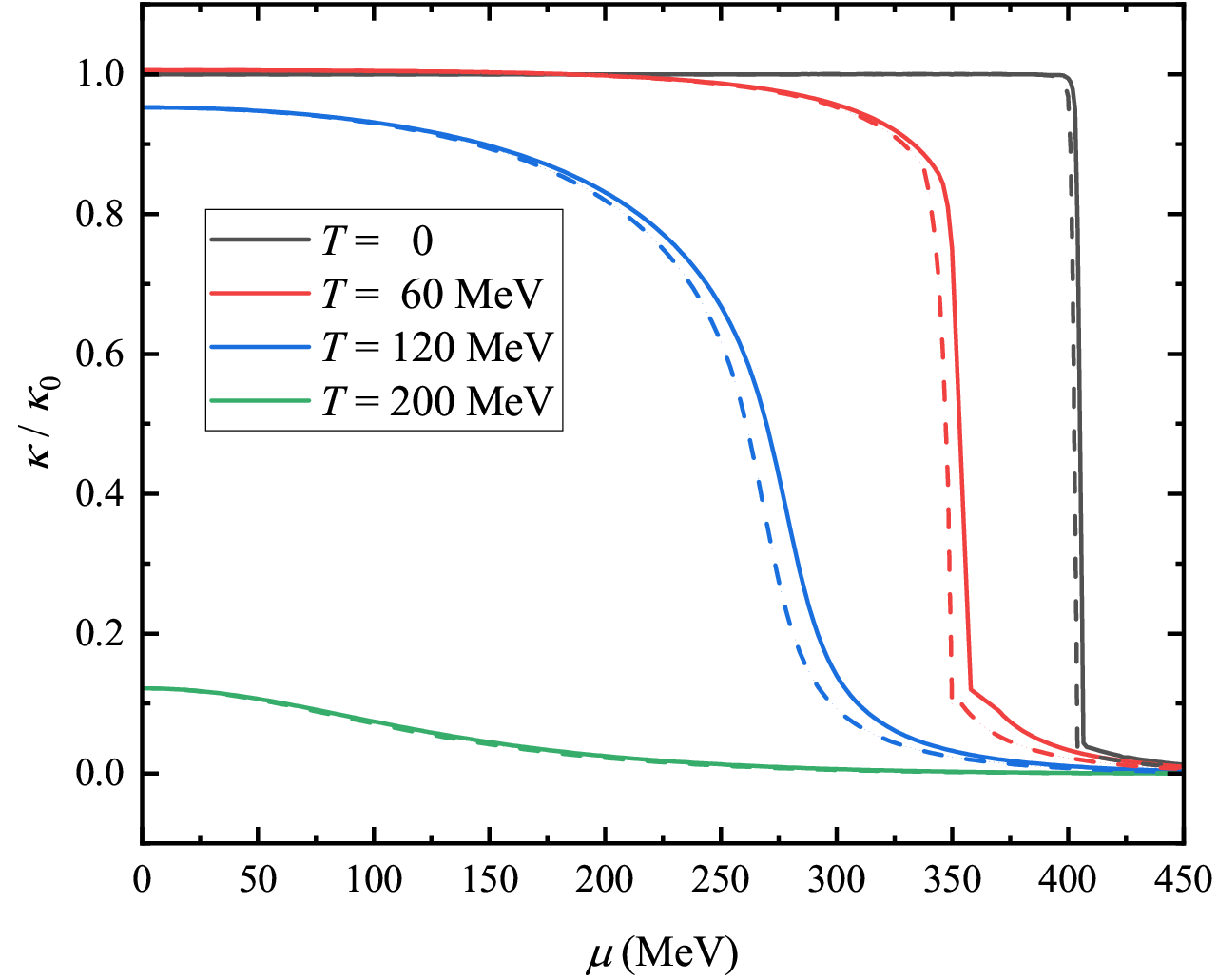}\\
  \caption{Domain wall tension as a function of chemical potential at various temperatures, comparing results with (solid) and without (dashed) the electric charge neutrality condition. 
  %Conventions for colors  are the same used in Fig.~\ref{fig:sigmamu}. 
  }\label{fig:kk0mu}
\end{figure}

Domain wall structures are ubiquitous in nature~\cite{Chen:2024wpw}, and they also play a distinctive role in fundamental particle physics. In particular, in the QCD vacuum a domain wall is realized as a static and spatially localized field configuration in which the axion field, or equivalently the effective $\theta$-parameter, interpolates smoothly between adjacent minima of the $\theta$-vacuum. These solutions arise because the QCD vacuum is not unique but possesses a set of discrete, degenerate minima related by shifts of $\theta \to \theta + 2\pi n$. Domain walls are thus topologically stabilized configurations that connect neighboring vacua, reflecting the nontrivial topology of the QCD vacuum. Their key physical property is the tension, which quantifies the energy per unit area required to separate regions belonging to different vacuum sectors. For an axion field $a(x)$ with decay constant $f_a$ and effective potential $V(\theta)=\Omega(\theta)-\Omega(0)$, where $\theta = a/f_a$ represents the dynamical QCD vacuum angle, the tension can be written as~\cite{Zhang:2023lij}
\begin{align}
\kappa = 2f_a\int_0^\pi d\theta\sqrt{2\,[V(\theta)-V(0)]}.
\end{align}

Figure~\ref{fig:kk0mu} illustrates the normalized domain wall tension, $\kappa/\kappa_0$, as a function of the chemical potential for various representative temperatures. At $T=0$, the tension remains essentially constant over a wide range of the chemical potential before dropping abruptly at the critical value, a clear signal of the first-order chiral phase transition. As the temperature increases, the onset of this drop shifts to smaller chemical potentials and the transition becomes progressively smoother, reflecting the weakening of the first-order character and the emergence of crossover behavior. At $T=200$ MeV, the tension is strongly suppressed across the entire range of chemical potential, indicating that thermal fluctuations substantially reduce the free-energy barrier separating distinct $\theta$ vacua. This reduction can be understood from the finite-temperature effective potential, where thermal excitations flatten the vacuum structure by lowering the depth of the minima and diminishing the barrier height. Consequently, the stability of axion domain walls is significantly reduced in hot and dense matter, with important implications for their dynamics in compact stellar environments~\cite{Huang:2024nbd}.

%%%%%%%%%%%%%%%%%
%\subsection{Chiral condensate and topological susceptibility}
\subsection{Topological susceptibility and the fourth cumulant}

\begin{figure}[htb]  %bt
   \includegraphics[width=0.48\textwidth]{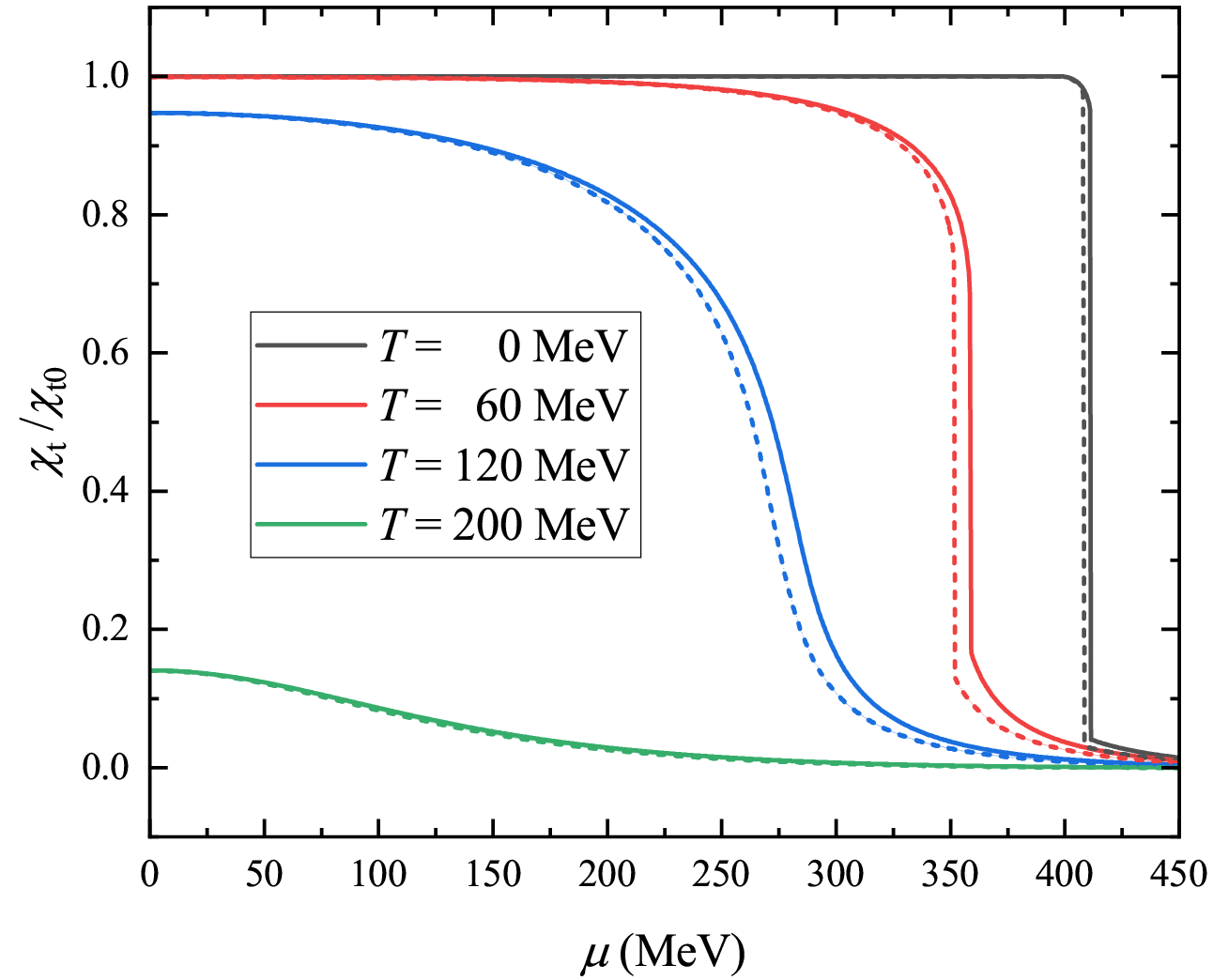}\\
  \caption{Topological susceptibility, scaled by its value in the vacuum, as a function of the temperature at different temperatures with (solid) and without (dashed) the charge neutrality condition. %Conventions for colors  are the same used in Fig.~\ref{fig:sigmamu}.
  }\label{fig:chimu}
\end{figure}

Knowing the $\theta$ dependence of the effective potential of the QCD vacuum makes it possible to evaluate the topological susceptibility and the higher-order cumulants as functions of the chemical potential.
The topological susceptibility is a fundamental quantity in QCD and plays a central role in the dynamics of the $U(1)_A$ channel~\cite{Cui:2022vsr}. It is defined as 
\begin{eqnarray} \label{eq:topological}
\chi_t=\frac{d^{2} V(\theta)}{d \theta^{2}}\bigg|_{\theta=0} .
\end{eqnarray}
In principle, this qunatity can be determined from lattice simulations, particularly at zero and low baryon chemical potentials. Here we focus on its evaluation within the NJL model. According to Eq.~(\ref{eq:topological}), the topological susceptibility in the physical vacuum at zero temperature is obtained as
$
\chi_t^{1/4}=79.87~\mathrm{MeV} ,
$
which is in good agreement with the CHPT estimate of Ref.~\cite{GrillidiCortona:2015jxo}, $\chi_t^{1/4}=77.8(4)$ MeV, as well as with lattice QCD result~\cite{Borsanyi:2016ksw}, $\chi_t^{1/4}=78.1(2)$ MeV, all evaluated under isospin symmetry.

In Fig.~\ref{fig:chimu}, we plot the topological susceptibility, scaled by its vacuum value, as a function of the chemical potential for several representative temperatures. At low temperatures such as $T=0$ and $T=60$ MeV, $\chi_t$ stays close to its vacuum magnitude over a wide range of chemical potential and then exhibits a sudden drop near the critical point, reflecting the first-order nature of the chiral phase transition and the abrupt restoration of chiral symmetry. At $T=120$ MeV, the decrease of $\chi_t$ becomes noticeably smoother, signaling the weakening of the first-order transition. At even higher temperature, $T=200$ MeV, the susceptibility decreases monotonically without a sharp discontinuity, characteristic of a smooth crossover regime. A comparison of the solid and dashed curves reveals that the enforcement of charge neutrality and $\beta$-equilibrium shifts the critical chemical potential slightly but leaves the qualitative features essentially unchanged. This robustness highlights that the temperature and density dependence of $\chi_t$ captures the essential interplay between QCD topology and chiral symmetry restoration, with direct implications for the behavior of axions and domain walls in hot and dense stellar matter.

\begin{figure}[htb] 
   \includegraphics[width=0.48\textwidth]{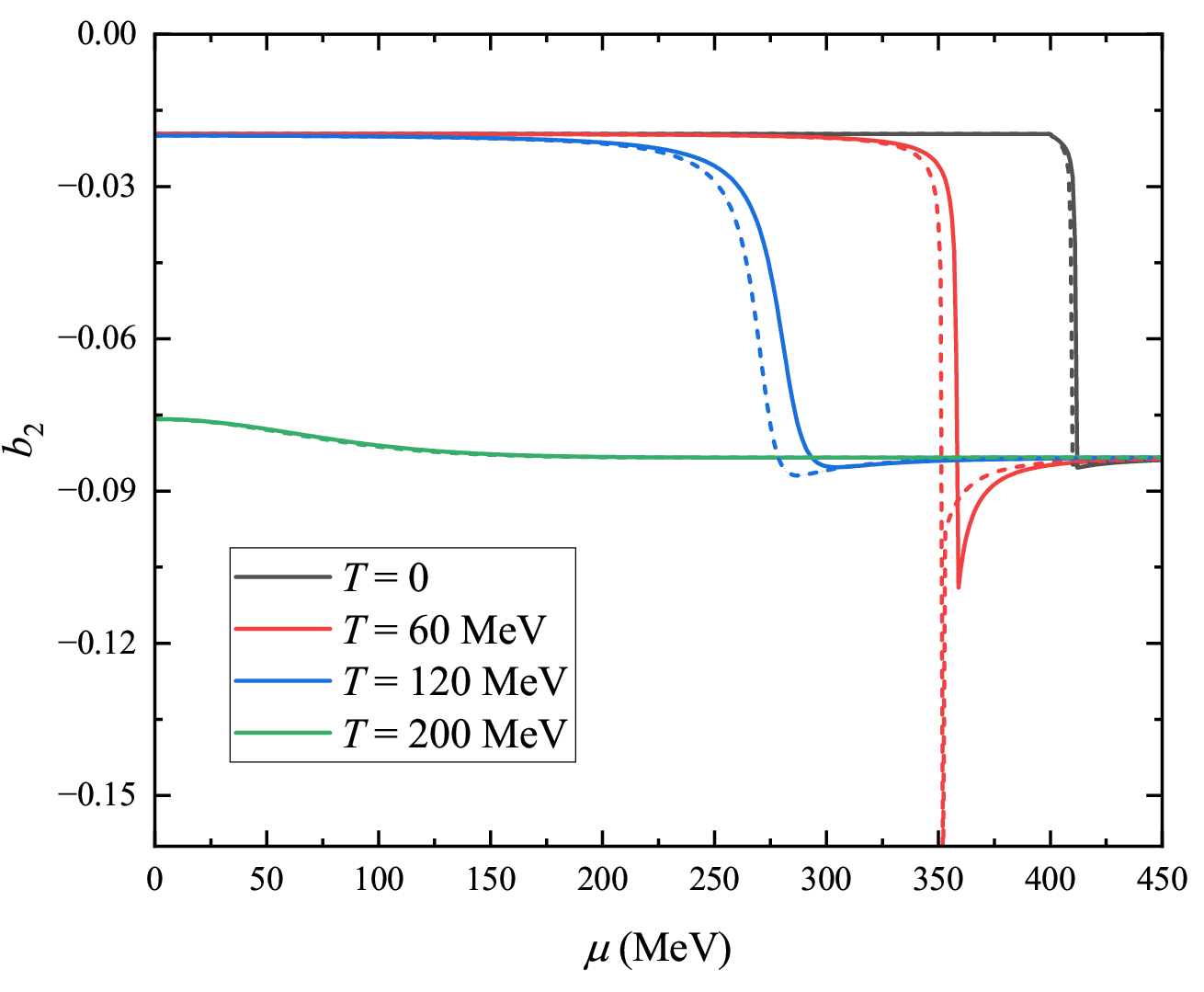}\\
  \caption{Normalized fourth cumulant, scaled by its value in the vacuum, as a function of the chemical potential at different temperatures with (solid) and without (dashed) the charge neutrality condition. 
  %Conventions for colors  are the same used in Fig.~\ref{fig:sigmamu}. 
  }\label{fig:b2Tmu}
\end{figure}

Figure \ref{fig:b2Tmu} displays the normalized fourth cumulant $b_{2}$, scaled by its vacuum value, as a function of the chemical potential at several representative temperatures. At zero and low temperature, $b_{2}$ remains almost constant at small chemical potentials and then exhibits a sharp discontinuity near the critical point, signaling the first-order nature of the chiral phase transition and the abrupt change of the axion effective potential around $\theta=\pi$. At intermediate temperature, exemplified by $T=120$ MeV, the variation of $b_{2}$ becomes smoother, indicating the weakening of the first-order transition. At higher temperature, $T=200$ MeV, the cumulant decreases gradually with increasing chemical potential without showing a distinct discontinuity, consistent with the expected crossover behavior.

The sign and magnitude of $b_{2}$ encode valuable information about the shape of the axion potential beyond the quadratic approximation. Negative values of $b_{2}$ imply stronger anharmonic corrections, which directly influence axion self-interactions. The pronounced dip of $b_{2}$ in the vicinity of the transition, particularly visible at $T=60$ MeV, highlights the emergence of non-Gaussian fluctuations in the topological charge distribution. A comparison between solid and dashed curves shows that imposing charge neutrality and $\beta$-equilibrium shifts the transition point only slightly while leaving the qualitative features unchanged. These findings underscore that the fourth cumulant is a sensitive probe of the QCD phase structure and plays an essential role in determining axion dynamics in hot and dense matter. 

\subsection{Axion mass and its self-coupling constant}

We recall that the topological susceptibility $\chi_t$ is directly connected to the axion potential and governs its temperature dependence~\cite{Borsanyi:2015cka}. The axion mass, which constitutes a fundamental parameter in this framework, is defined through the curvature of the effective potential $\mathcal{V}(a)$ at the minimum. Explicitly, it is given by the second derivative of $\mathcal{V}(a)$ with respect to the axion field $a$, evaluated at $a=0$,
\begin{eqnarray} \label{eq:ma2}
m_a^2=\left.\frac{\mathrm{d}^2 \mathcal{V}(a)}{\mathrm{d} a^2}\right|_{a=0}=\frac{\chi_t}{f_a^2},
\end{eqnarray}
where $\chi_t$ denotes the topological susceptibility. In addition, the axion self-coupling $\lambda_a$, which characterizes the strength of axion–axion interactions, is defined through the fourth derivative of the effective potential evaluated at $a=0$, 
\begin{eqnarray}
\lambda_a=\left.\frac{\mathrm{d}^4 \mathcal{V}(a)}{\mathrm{d} a^4}\right|_{a=0} .
\end{eqnarray}

In this subsection, we analyze the axion mass and self-coupling within the NJL model at finite temperature. The axion mass is of particular importance in experimental searches, most notably in cavity microwave experiments where axions are expected to convert into photons in the presence of a strong magnetic field~\cite{Bradley:2003kg,Kim:2008hd,Marsh:2015xka}. Within the NJL model, the axion mass at zero temperature is defined by Eq.~(\ref{eq:ma2}), which provides the baseline for extending the analysis to finite-temperature conditions. Numerically, we obtain
$%\begin{eqnarray}
m_af_a=6.38\times10^3~\mathrm{MeV}^2,
$%\end{eqnarray}
which is consistent with the prediction of CHPT, $m_af_a=6.06(5)\times10^3~\mathrm{MeV}^2$ in the isospin-symmetric limit, as well as with the standard invisible axion model estimate, $m_af_a\simeq6.0\times10^3~\mathrm{MeV}^2$~\cite{Kim:1986ax,Cheng:1987gp,Turner:1989vc,Raffelt:1990yz}.
The axion self-coupling is another quantity of interest, as it may play a role in the formation and stability of axion stars~\cite{Braaten:2015eeu,Bai:2016wpg}. At zero temperature, the NJL model yields 
$%\begin{eqnarray}
\lambda_af_a^4=-(55.64~\mathrm{MeV})^4,
$%\end{eqnarray}
in excellent agreement with the CHPT result $\lambda_af_a^4=-(55.79(92)~\mathrm{MeV})^4$~\cite{GrillidiCortona:2015jxo} for the case of two degenerate quark flavors.

\begin{figure}[htb]  %bt
  \includegraphics[width=0.48\textwidth]{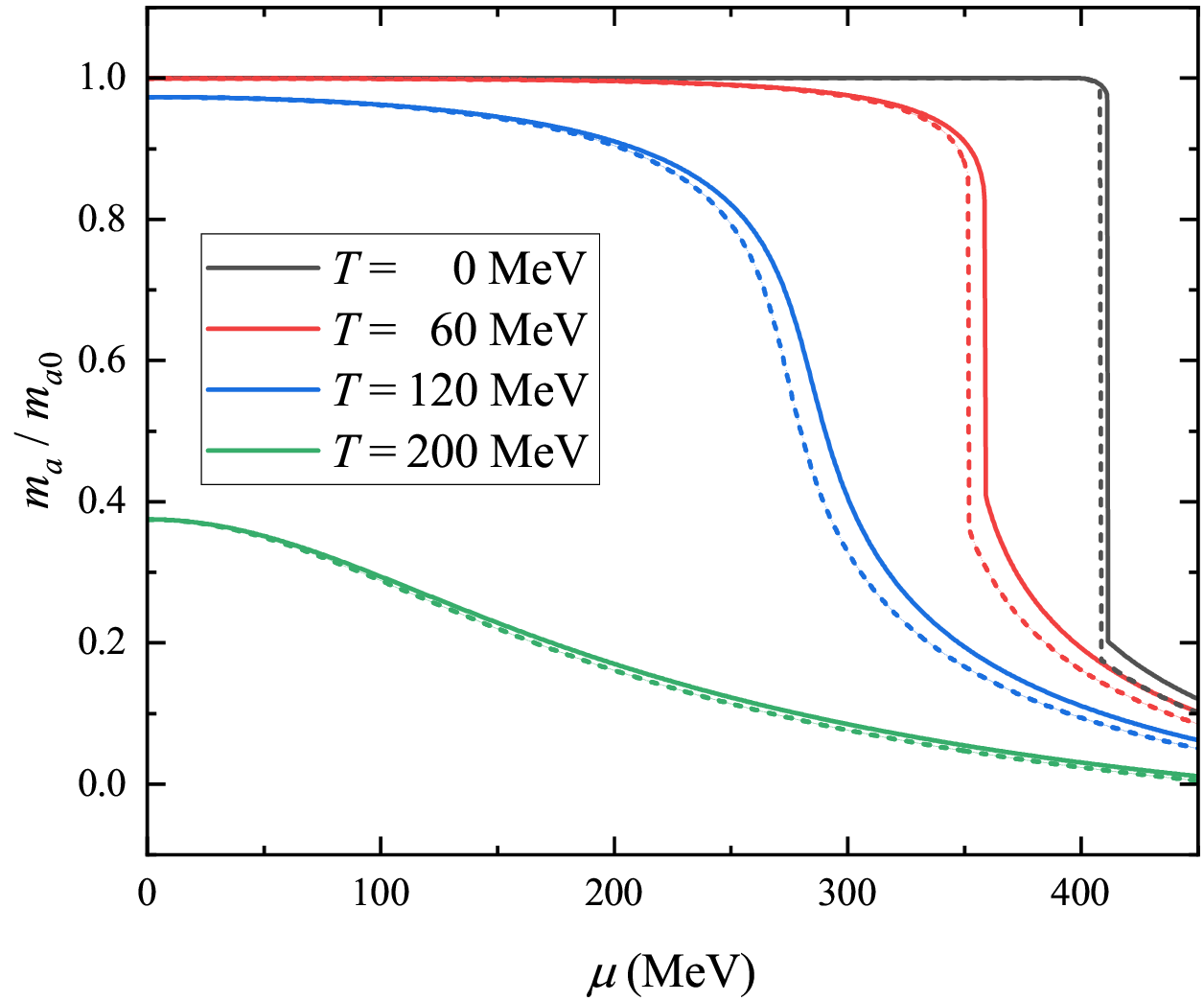}
  \caption{Axion mass, scaled by its value in the vacuum, as a function of the chemical potential at different temperatures with (solid) and without (dashed) the charge neutrality condition. %Conventions for colors  are the same used in Fig.~\ref{fig:sigmamu}.
  }\label{fig:axionmass}
\end{figure}

As shown in Fig.~\ref{fig:axionmass}, we plot the axion mass scaled by its vacuum value as a function of the chemical potential for several selected temperatures, both with and without the charge neutrality condition.  
These results have been obtained by using Eq.~(\ref{eq:ma2}) with 
the solution of the gap equation from Eq.~(\ref{eq:gapEquations2}).   The overall trend closely parallels that of the chiral condensate: at low temperature the mass remains nearly unchanged over a wide range of chemical potential and then exhibits a rapid suppression near the chiral transition, while at higher temperature the decrease becomes smoother and extends over a broader interval. This correlation reflects the intimate connection between the axion mass and the QCD topological susceptibility, both of which are governed by the chiral order parameter. The comparison between solid and dashed curves further shows that enforcing charge neutrality shifts the position of the transition but does not alter the qualitative pattern, underscoring that the dominant physics is driven by the critical behavior of the underlying chiral dynamics.

\begin{figure}[htb]  %bt
   \includegraphics[width=0.48\textwidth]{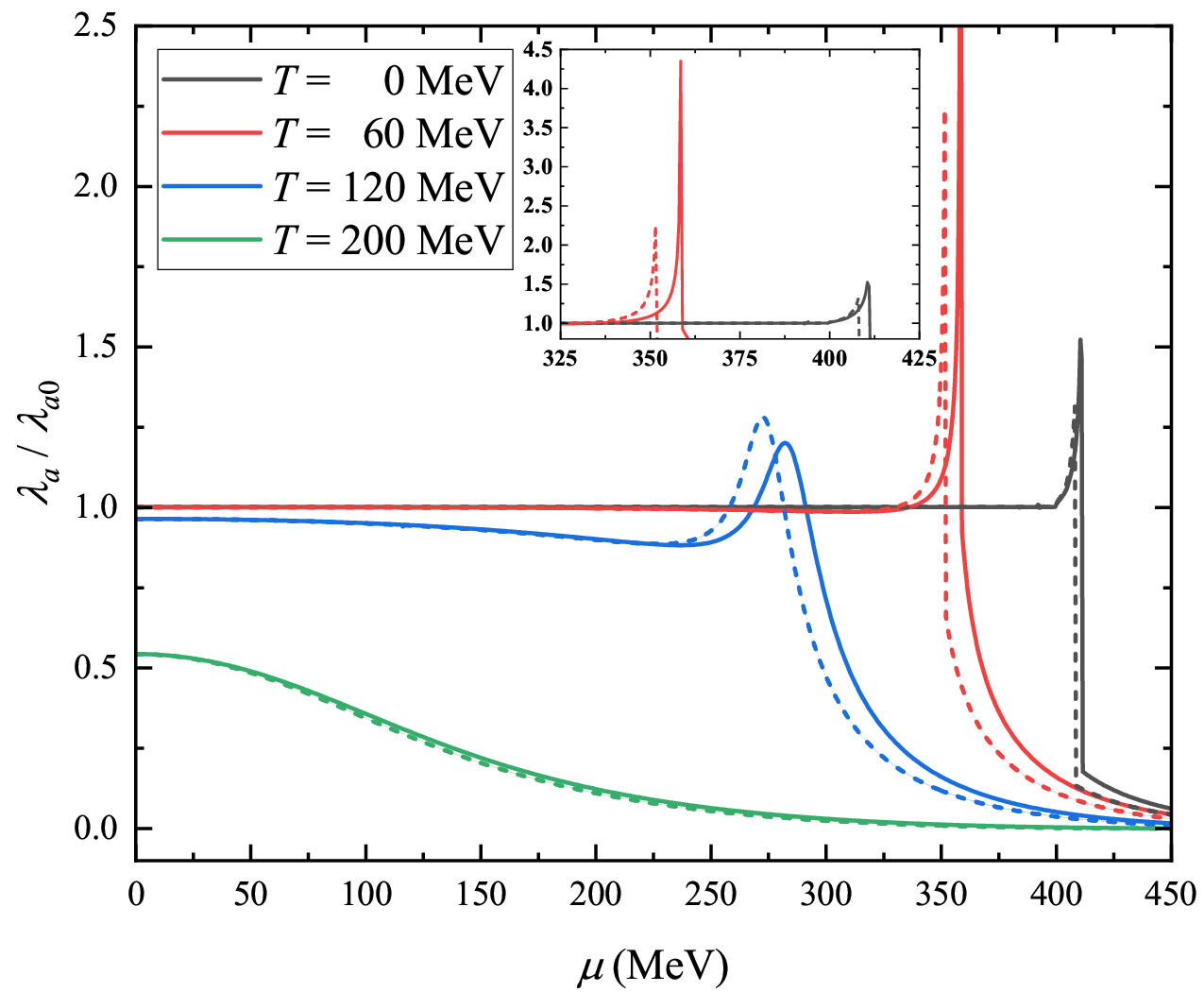}\\
  \caption{Axion self-coupling constant, scaled by its value in the vacuum, as a function of the chemical potential at different temperatures with (solid) and without (dashed) the charge neutrality condition. 
  %Conventions for colors  are the same used in Fig.~\ref{fig:sigmamu}. 
  }\label{fig:lambdafamu}
\end{figure}

Figure \ref{fig:lambdafamu} presents the axion self-coupling constant, scaled by its vacuum value, as a function of the chemical potential at several representative temperatures. At low temperatures such as 0 and 60 MeV, the self-coupling stays nearly constant at small chemical potentials and then shows a sharp and pronounced peak close to the critical chemical potential, reflecting the strong enhancement of nonlinear axion interactions at the first-order chiral phase transition. At $T=120$ MeV the peak is still visible but becomes broader and less pronounced, while at $T=200$ MeV the coupling decreases smoothly with increasing chemical potential without a singular structure, consistent with a crossover regime. The comparison between the solid and dashed curves shows that enforcing charge neutrality and beta equilibrium slightly shifts the position of the critical peak but does not change its overall shape or magnitude, indicating that the dramatic enhancement of the self-coupling near the phase boundary is a robust feature of dense QCD matter. Since the axion self-coupling is determined by higher-order derivatives of the QCD effective potential, its critical behavior encodes the non-Gaussian nature of topological charge fluctuations. The large amplification observed near the transition, which can reach several times the vacuum value, implies that axion self-interactions in compact stars may be substantially stronger than in vacuum, with potential consequences for the stability of axion domain walls and the dynamics of dense stellar matter.

%------------------------------------
\begin{table}
%\squeezetable
\caption{\label{table:lambda} 
Maximum values of the normalized axion self-coupling constant, $\lambda_a/\lambda_{a0}$, at fixed temperatures, obtained under the charge neutrality constraint. It can be seen that near the phase transition point, the axion self-coupling constant shows a remarkable enhancement effect, with a maximum value that can reach seven times its vacuum value.}
\setlength{\tabcolsep}{0.9pt}
\renewcommand\arraystretch{1.7}
\begin{ruledtabular}
\vspace{+0.1cm}
\begin{tabular*}{\hsize}{@{}@{\extracolsep{\fill}}ccc@{}}
$T$ (MeV)      &  $\mu$ (MeV)      &   $\lambda_a/\lambda_{a0}$   \\
      %   & fm  &   (MeV)    &   (MeV)  &  (MeV)  &  (MeV)  &  Msun & (km )     \\   
         \hline
0     &  436.0       &  3.64    \\
20    &  401.5       &  1.71 \\
40    &  382.0       &  2.56    \\
60     &  358.5       &  4.35    \\
70     &  346.5       &  7.34    \\
80      &  335.0       &  6.55   \\
100      &  311.5      &  1.77   \\
120      &  282.0       &  1.20   \\
\end{tabular*}
\end{ruledtabular}
\vspace{-0.5cm}
\end{table}
%------------------------------------

Table~\ref{table:lambda} summarizes the maximum values of the axion self-coupling constant at several fixed temperatures under the charge neutrality condition. The results show that the enhancement of the coupling strongly depends on both temperature and chemical potential. At zero temperature the maximum reaches about $3.6$ times the vacuum value, while at $T=70$ MeV it rises dramatically to more than seven times the vacuum value, reflecting the critical enhancement near the chiral phase transition. As the temperature increases further, the peak value decreases, approaching nearly the vacuum level at $T=120$ MeV. This non-monotonic behavior highlights the intimate connection between axion self-interactions and the underlying QCD phase structure. The large amplification of the self-coupling near the critical region implies that axion dynamics in dense stellar matter could be markedly different from those in vacuum, potentially affecting the stability of axion domain walls and the transport properties of compact stars. Such strong self-interactions may also leave imprints on the cooling history and oscillation modes of compact stars, thereby offering a possible observational window into axion physics.

%%%%%%%%%%%%%%%%%%%%%%%%%%%%%%%%%%%%%%%%
\section{Conclusions}  \label{sec:CONCLUSION}

In this work, we have investigated the behavior of QCD axions and domain walls in hot and dense matter within the NJL model, enforcing both charge neutrality and $\beta$-equilibrium to model compact stellar matter realistically. Our results reveal several novel features. We find that the axion mass is strongly suppressed near the chiral restored phase, while the self-coupling constant develops a pronounced enhancement that can exceed its vacuum value by more than a factor of seven around the critical region. This sharp amplification had not been quantified in earlier finite-temperature studies and highlights the intimate link between the QCD phase structure and in-medium axion interactions. Moreover, by analyzing the thermodynamic potential and higher-order cumulants of the topological charge distribution, we demonstrate that axion properties are highly sensitive not only to the location but also to the nature of the phase transition. In particular, the domain wall tension decreases significantly with both temperature and chemical potential, indicating a destabilization of axion domain walls in dense environments.

These findings advance our understanding in two respects. Theoretically, they extend the NJL model as a tool for investigating the $\theta$ vacuum by capturing the interplay of scalar, pseudoscalar, and diquark condensates under neutrality and equilibrium constraints, with an explicit emphasis on density evolution. Phenomenologically, they suggest that axion self-interactions can become strongly enhanced and domain walls destabilized in the cores of neutron stars or quark stars, potentially influencing cooling, transport, and the dynamics of supernovae or merger events. 
While our analysis provides the first systematic study of these effects from the perspective of density evolution under realistic constraints, future efforts should refine the framework by incorporating vector and axial-vector channels, benchmarking against lattice QCD, and extending to include hadronic degrees of freedom. Such developments will be essential for establishing whether axion-induced modifications to the QCD phase structure can manifest as observable astrophysical signatures.

\section*{Acknowledgments}

This work is supported by  
the National Natural Science Foundation of China 
(Grant Nos.~12205093, 12404240, 12405054, and 12375045), and the Hunan Provincial Natural Science Foundation of China (Grant Nos.~2021JJ40188 and 2024JJ6210). M.R. also acknowledges support from the European Union – Next Generation EU through the research grants number P2022Z4P4B “SOPHYA - Sustainable Optimised PHYsics Algorithms: fundamental physics to build an advanced society”, under the program PRIN 2022 PNRR
of the Italian Ministero dell’Universit\`a e Ricerca (MUR).

\bibliographystyle{aapmrev4-2}  %%y%%  %% longbibliography
\bibliography{Ref}

%\end{CJK*}
\end{document}